\documentclass[12pt,a4paper]{article}
\usepackage{amsmath,amssymb,mathrsfs}
\usepackage{graphicx,epsfig}
\usepackage{color}
\newcommand{\GF}{G_{{\rm F}}}

\newcommand{\GeV}{\,{\rm GeV}}
\newcommand{\TeV}{\,{\rm TeV}}

\newcommand{\eq}[1]{(\ref{eq:#1})}

\newcommand{\mst}{m_{\tilde t}}
\newcommand{\lag}{\mathscr{L}}
\newcommand{\vev}[1]{\langle #1 \rangle}
\newcommand{\abs}[1]{\left\vert#1\right\vert}

\newcommand{\hc}{\mbox{{\rm h.\,c.}}}

\begin{document}
\baselineskip 0.6cm

\begin{titlepage}

\begin{flushright}
\end{flushright}

\vskip 1.0cm

\begin{center}
  {\Large \bf Relating the ElectroWeak scale to an extra dimension:
  constraints from ElectroWeak Precision Tests}

  \vskip 1.0cm

  {\large Guido Marandella, Michele Papucci}

  \vskip 0.5cm

  {\it Scuola Normale Superiore and INFN, Piazza dei Cavalieri 7,
    I-56126 Pisa, Italy} \\

  \vskip 1.0cm

  \abstract{In models with supersymmetry breaking by boundary
  conditions on an extra dimension it is possible to relate in a
  quantitative natural way the Fermi scale to the size of the extra
  dimension. We analyze in detail
  the compatibility of such models with ElectroWeak
  Precision Tests.}

\end{center}
\end{titlepage}

\section{Introduction}
\label{sec:intro}

In the last decade the Standard Model (SM) has received impressive
experimental confirmations by the ElectroWeak Precision Tests (EWPT).
However, despite its experimental success, the core of the SM,
i.e. the physical mechanism of ElectroWeak Symmetry Breaking (EWSB),
remains unexplained. The SM, through the Higgs mechanism,
probably catches the essence of EWSB, without providing a physical theory for
it. The Higgs boson mass has a quadratic sensitivity to the
ultraviolet (UV) physics. This fact led to the expectation of
discovering New
Physics around the TeV scale. Today, thanks to EWPT, the problem of
the stabilization of the Higgs potential assumes a new, entirely low energy, aspect. 
The 1-loop correction to the Higgs boson mass due to the top quark is
given by
\begin{equation}
  \label{eq:higgs1looptop}
  \delta m_H^2 = \frac{3}{\sqrt{2} \pi^2} \GF m_t^2 \Lambda^2 \simeq
  (120 \GeV)^2 \; \left( \frac{\Lambda}{430 \GeV} \right)^2
\end{equation}
The EWPT suggest an indirect evidence for a light Higgs boson ($m_H<237 \GeV$ at 95 \% of
C.L. \cite{EWWG}). Without invoking a very finely-tuned cancellation between a
counterterm and the loop correction \eq{higgs1looptop}, one requires
the  latter not being
much larger than the physical Higgs mass. One then needs a physical
mechanism, such as the appearance of some new particle, which cuts off
$\delta m_H^2$ not too far from the ElectroWeak scale. The fact
that the EWPT showed no signals of physics beyond the
SM makes the gap between the Higgs mass and a typical scale of new
physics problematic from the naturalness point of view, even if this
scale is in the $5\div 10 \TeV$ range. This
is the so-called ``Little Hierarchy Problem''. 

These kind of arguments led many to the expectation of seeing some
supersymmetric particle at LEP. But this did not
happen. In Supersymmetry it is the appearance of a stop quark with
mass $\mst$ which
cuts off eq. \eq{higgs1looptop} rendering logarithmic the quadratic
divergence. However, the fact that neither the Higgs boson nor any
supersymmetric particle has been
experimentally discovered implies, in the context of supersymmetric
models, that a certain amount of fine-tuning has
necessarily to be accepted.

All this motivates further thoughts on the problem of EWSB. Precisely,
one should look for alternative models which i) are possibly predictive,
ii) naturally foresee the existence of a light (not fine-tuned)
Higgs boson and iii) are not in conflict with EWPT. We want to focus
on a recent proposal of supersymmetry breaking obtained by imposing
boundary conditions along a compact extra dimension
\cite{Barbieri:2000vh,Barbieri:2002uk,Barbieri:2002sw}. This approach
allows to connect the ElectroWeak scale to the radius of
compactification of an extra dimension. The implementations
of this idea reach the goals i) and ii). In fact the models which have
been built are extremely predictive in terms of a minimal number of
free parameters and the connection between the Higgs mass and the
radius of compactification can be established with very high precision
\cite{Barbieri:2003kn}. The ``Little Hierarchy problem'' is solved
since the Higgs mass is UV-insensitive. In
this paper we want to analyze in detail the point iii): the compatibility of these
models with EWPT.

The paper is organized as follows.  Mostly for the ease of
the reader, we recall in Sec. \ref{sec:model} the main features of the
idea to connect the ElectroWeak scale
to the radius of compactification of an extra dimension and describe
the two possible implementations of this setup
(Subsec. \ref{sec:CSM} and \ref{sec:QLT}). In Sec. \ref{sec:EWPT} we
analyze in detail the impact of EWPT on the two implementations. The
conclusions are drawn in Sec. \ref{sec:conclusions}.

\section{The models}
\label{sec:model}

\subsection{The setup}
\label{sec:setup}

\begin{figure}[t!]
  \centering
  \includegraphics[width=9cm]{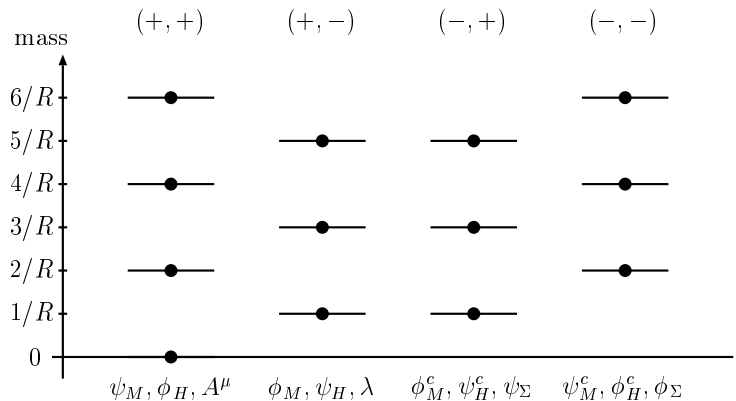}
  \caption{Tree-level KK mass spectrum of a multiplet (vector,
matter or Higgs) with the indicated boundary conditions.}
  \label{fig:spectrum}
\end{figure}

Let us consider a $SU(3) \times SU(2) \times U(1)$ invariant theory
with a compact spatial extra-dimension, seen in the following as a segment of
length $\pi R/2$. We require that the 5D theory is
supersymmetric. Given a generic  field $f$, supersymmetry implies the
existence of its superpartner $\tilde f$. Furthermore the existence of the conjugated fields
$f^c,\tilde f^c$ follows from
5D Poincar\'e invariance. They have the same chirality as $f,\tilde f $ but
opposite quantum numbers. 
Up to now every field has a zero mode. It is possible, in a
unique way, to get rid of the unwanted states and remain with only the
SM zero modes. This can be achieved by imposing suitable boundary
conditions to the fields of the theory. In fact imposing 
Dirichlet $(+)$ or Neumann $(-)$ conditions at the two
boundaries, only $(+,+)$ fields have a zero mode:  $(+,-)$ and $(-,+)$
lightest modes are at $1/R$, while the $(-,-)$ are at $2/R$. The assignment of
the boundary conditions to the various fields is shown in
Fig. \ref{fig:spectrum}. 

Now the mass of every particle is shifted
of $1/R$ with respect to its superpartner: the boundary conditions
have non-locally broken 5D supersymmetry. This mechanism is known as
supersymmetry breaking $\grave{\textrm{a}}$ la Scherk-Schwarz
\cite{Scherk:1979ta,Scherk:1979zr}. However there is a residual
local supersymmetry. In fact the initial \mbox{$N=1$} 5D supersymmetry can be seen,
from the 4D point of view, as two \mbox{$N=1$}, 4D supersymmetries
$\xi_1,\xi_2$ linked by an $SU(2)_R$ symmetry.
After imposing the boundary conditions the theory is still
supersymmetric under
$(\xi_1^{(+,-)},\xi_2^{(-,+)})$. The two supersymmetric transformation
parameters have thus to satisfy proper boundary conditions.
Then at the
boundaries there are two different $N=1$, 4D supersymmetries: at $y=0$
one has $(\xi_1 \neq 0, \; \xi_2 = 0)$; at $y=\pi R/2$ one has $(\xi_1
= 0, \; \xi_2 \neq 0)$.
This residual supersymmetry
is crucial for the calculability of the model: it renders
insensible to the UV physics several observables.

Furthermore, after imposing the boundary conditions, there are two
other residual symmetries.
\begin{itemize}
\item A continuous R-symmetry with R-charges given in Table
\ref{tab:r-charges}, intact even after EWSB. The absence of any
A-terms or Majorana gaugino masses can be traced back to this
symmetry.

\item A local $y$--parity $\mathcal{P}_5$ under which any field transforms as
\begin{equation}
\varphi (y) \rightarrow \eta \, \varphi (\pi R /2 -y)
\label{eq:y-parity}
\end{equation}
where $\eta$ is the parity assignment at any one of the two
boundaries.
This symmetry forbids
local mass terms for the hypermultiplets.
\end{itemize}

\begin{table}
\begin{center}
\begin{tabular}{|c||c|c|c|} \hline
$R$  & gauge $V$           & Higgs $H$     & matter $M$
\\ \hline
+2   &                     & $h^c$         &                          \\
+1   & $\tilde{\lambda} $  & $\tilde{h}^c$ & $\tilde{m}, \tilde{m}^c$ \\
0    & $A^\mu, A^c$        & $h$           & $m, m^c$                 \\
$-1$ & $\tilde{\lambda}^c$ & $\tilde{h}$   &
\\ \hline
\end{tabular}
\caption{Continuous $R$ charges for gauge, Higgs and matter
components. Here, $m$ represents $q, u, d, l, e$.}
\label{tab:r-charges}
\end{center}
\end{table}

The most general Lagrangian compatible with the symmetries of the theory is the following
\begin{equation}
  \label{eq:lagrangian}
  \lag = \lag_5 \; + \; \delta(y) \; \lag_4 \; + \; \delta(y-\frac{\pi R}{2}) \; \lag'_4
\end{equation}
where $\lag_5$ is supersymmetric under the full $N=1$, 5D
supersymmetry, while $\lag_4$ and $\lag'_4$ respect the two different
$N=1$, 4D supersymmetries at the boundaries. The Yukawa interactions,
necessary to give a mass to the SM fermions, are forbidden by 5D
supersymmetry and thus have to be localized at the
boundaries. We take the top quark Yukawa coupling, crucial for EWSB,
to be localized at $y=0$:
\begin{equation}
  \label{eq:yukawa}
   \delta (y)  \int d^2 \theta \; \lambda_U \;  Q U H  
+ {\rm h.c.} 
\end{equation}
while the bottom Yukawa coupling is necessarily localized at the opposite
boundary. The tree-level Higgs potential is constrained from
supersymmetry to have the form 
\begin{equation}
  \label{eq:treepotential}
  V_{H,0}=\frac{g^2+g'^2}{8} \abs{\phi_H}^4
\end{equation}
where $g$ and $g'$ are the 4D $SU(2)$ and $U(1)$ gauge couplings.

There are two possible implementations of this setup. The first one
is the so-called ``Constrained Standard Model'' (CSM)
\cite{Barbieri:2000vh,Barbieri:2002uk} and the second
one is the so-called ``Quasi-Localized Top Model'' (QLTM)
\cite{Barbieri:2002sw,Barbieri:2003kn}. We now
describe them in more detail.

\subsection{The CSM}
\label{sec:CSM}

If we implement the setup previously exposed with no modification, we
are in the position to compute the 1-loop Higgs boson mass  due to
the top Yukawa coupling \eq{yukawa} and, more in
general, the Coleman-Weinberg potential in order to analyze EWSB. The
1-loop Higgs mass is given by
\begin{equation}
  \delta m_H^2 = - \frac{63 \, \zeta (3)}{\sqrt{2} \pi^4} \frac{G_F \;
  m_t^2}{R^2} \simeq - \frac{0.19}{R^2}  \label{eq:corr-m2-result}
\end{equation}
The finiteness of
(\ref{eq:corr-m2-result}) is a consequence of local supersymmetry
conservation in 5D, as previously discussed. Thus EWSB is triggered
radiatively by the top Yukawa interaction.

One can then compute the full 1-loop potential due to the top
multiplet. It is a function of the top Yukawa coupling (determined by
the top quark mass) and of $1/R$: there is one parameter less then in
the SM, hence the name of the model. Requiring that the Higgs potential has its minimum at the Higgs
vev $v$, determined by the Fermi constant $G_F$, one can compute the
value of $1/R$ which turns out to be in the $400 \GeV$ range
\cite{Barbieri:2000vh}. With
this value of $1/R$, taking the second derivative of the potential
evaluated at its minimum, one finds the Higgs mass, predicted to
be about $130 \GeV$ \cite{Barbieri:2000vh}.

In the CSM with a single Higgs supermultiplet a
quadratically divergent Fayet-Iliopoulos (FI) term arises for the
hypercharge vector supermultiplet
\cite{Ghilencea:2001bw}. However, given the value of the cut-off,
determined below in Sect. \ref{sec:cutoff}, the FI is numerically negligible \cite{Barbieri:2001cz}.
Nevertheless, in order to ensure the quantum consistency of the theory,
one has to give up the local parity \eq{y-parity}. Then it is
necessary to consider possible mass parameters for all the
hypermultiplets. If they are taken to be small with respect to $1/R$
this consists of a small modification of the theory
\cite{Barbieri:2002uk}. If they are
$\gtrsim 1/R$ one has a qualitatively different model,
described in the next section.

\subsection{The QLTM}
\label{sec:QLT}

Giving up the local parity \eq{y-parity} one can consider the
effect of a mass term for every hypermultiplet. For the hypermultiplet of
components $(\psi,\psi^c,$ $\varphi,\varphi^c)$, the 5D mass term is
\begin{eqnarray}
\lag_m &=&-\left(\psi m(y) \psi^c + \hc \right)
- M^2 \left(\abs{\varphi}^2+\abs{\varphi^c}^2 \right) \nonumber\\
 && -2M \left(\delta(y) + \delta
(y-\pi R/2)\right)\left(\abs{\varphi}^2-\abs{\varphi^c}^2\right)
\label{eq:lagrangian-mass}
\end{eqnarray}
The mass term $m(y)$ must respect $(-,-)$ boundary conditions. We consider a constant mass term inside the segment
$(0,\pi R/2)$. For a matter hypermultiplet all the KK modes acquire a mass
which goes as $\abs{M}$ for $\abs{MR} \gg 1$. The fermionic zero mode remains
massless, while one of the two lightest scalars has mass which goes to
zero as $\abs{MR} \to \infty$. The wave function of the fermion zero mode
gets exponentially localized at on boundary depending on the sign of
$M$
\begin{equation}
  \label{eq:wavefunction}
  \psi_0 (y) \propto \abs{M}^{1/2} \exp \left[-M y+ (M-\abs{M}) \pi R/2 \right]
\end{equation}
The same thing happens for the scalar particle which becomes
massless. In the limit $MR \to \infty$ a supersymmetric spectrum is
recovered.

\begin{figure}[t]
  \centering
  \includegraphics[width=12cm]{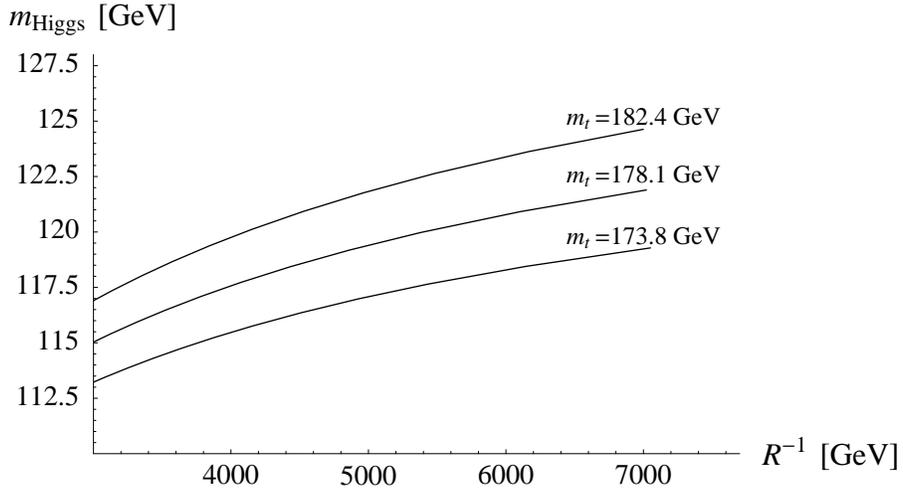}
  \caption{The Higgs function as a function of $1/R$ for $M_U=M_Q=M_D$
  and $M_H=0$.}
  \label{fig:MHvs1/R}
\end{figure}

We are now interested in the following model. We introduce a
second Higgs hypermultiplet $H_d$ in order to exactly cancel the 1-loop generated FI term. This additional Higgs doublet does not get a
vev and is irrelevant for EWSB \cite{Barbieri:2002sw}. The only mass
terms which are relevant for EWSB
are those for the third quark generation and for the Higgs. We
choose to localize the third generation of quarks towards the $y=0$
boundary through a common mass term $M_U=M_Q=M_D \equiv M$, setting
for the time being
$M_H=0$. With this configuration the FI term
vanishes.  Now one can compute
the 1-loop Higgs potential which now depends on two parameters:
$1/R$ and $MR$. Thus the Fermi constant allows to connect $MR$ with
$1/R$. Then the Higgs mass is a function of $1/R$. The connection
between the Higgs mass and $1/R$ can be established with a few percent
accuracy, calculating the relevant 2-loop contribution to the Higgs
potential
\cite{Barbieri:2003kn}, even if the relation between $1/R$ and $MR$ is
known with significantly less precision. We are interested in
typical values $MR \simeq 1.5 \div 2.5$: for too large values of $MR$ EWSB
does not occur successfully, while for too small values of $MR$ one
needs a mass term for the Higgs in order to stabilize the potential.
The result is shown in
\mbox{Fig. \ref{fig:MHvs1/R}}. The typical value of $1/R$ has increased of about
one order of magnitude with respect to CSM. This is due to the near
cancellation between the 1-loop electroweak correction and the top 2-loop correction to the Higgs
mass. 

\begin{figure}[t]
  \centering
  \includegraphics[width=12cm]{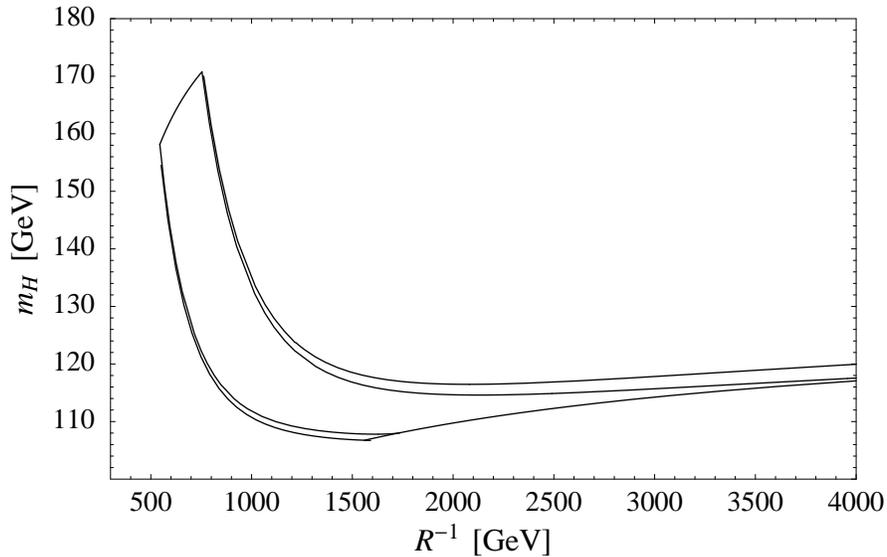}
  \caption{The Higgs function as a function of $1/R$ for $M_U=M_Q=M_D$
  and $M_H \neq 0$.}
  \label{fig:casomh}
\end{figure}

If one allows an equal mass term $M_H$ for the two Higgses the FI
still vanishes. This affects EWSB giving a tree level mass term to the
Higgs potential that stabilizes the Higgs potential, allowing lower
values of $MR$. In Fig. \ref{fig:casomh} we show the Higgs mass as a function
of $1/R$ leaving $M_H$ free to vary. However $M_H$ is limited from
above by the
request of a maximum cancellation of $10\%$ in the slope of the
potential and from below by the stability of the
potential. In the upper range of $1/R$, $M_H$ is irrelevant and we
recover the previous results. The
calculation of the potential done in \cite{Barbieri:2003kn} is
reliable either for $1/R \lesssim 1 \TeV$ (where the 1-loop top contribution
dominates) or for $1/R \gtrsim 2 \TeV$ (where a quasi-localized
approximation can be used). For intermediate values of $1/R$ a more
accurate calculation of the 2-loop contribution could be necessary.

\subsection{The UV cutoff}
\label{sec:cutoff}

The previous models are formulated in 5D and are
non-renormalizable. Indeed they have to be regarded as effective field
theories, valid up to an energy scale $\Lambda$ where they will be
completed by some kind of a more fundamental theory.

Since perturbative calculations in an effective field theory are
organized as an expansion series in $(E/\Lambda)$, the range of their
validity is controlled by the size of the cutoff $\Lambda$. It is
then necessary to correctly estimate its value.

A lower bound on the cutoff can be given by estimating the
scale at which one of the interactions of the model ceases to be
perturbative. Then the theory (or at least one of
its sectors) should be completed by UV unknown physics at energies
of the order of $\Lambda$.

As already shown in \cite{Barbieri:2000vh,Barbieri:2002sw} the first interactions that become
strong are the Yukawa couplings of the 3rd generation.

This is due to the fact that Yukawa couplings are localized at the
boundaries and break discrete 5th-momentum
conservation. The number of open channels in a process mediated by
a Yukawa interaction grows more than linearly with the energy, bringing
5D Yukawa couplings to become strong before any gauge
coupling of the same 4D strength.

Here, we recall the estimation of $\Lambda$ from top and bottom
Yukawa interactions. We consider also bottom interactions because in
presence of hypermultiplet masses the hierarchy $m_b/m_t$ can be
generated from localization effects alone and the size of $\lambda_t$ and
$\lambda_b$ in 5D can be comparable.

Let us focus first on the CSM case. Here only $\lambda_t$ matters since
all the hypermultiplets masses are zero. $\lambda_t(\Lambda)$ can
be estimated either by Naive Dimensional Analysis (NDA)
\cite{Manohar:1984md} properly adapted to 5D \cite{Chacko:1999hg} or by 4D NDA, taking into account the
number $N$ of KK modes below the cutoff.
In the first case we get
\begin{equation}
  \label{eq:cut-5D-NDA}
\widehat \lambda_t(\Lambda) \simeq \frac{1}{16 \, \pi^2} \left(\frac{24 \,
\pi^3}{2 \, \pi \, \Lambda R} \right)^{3/2} \simeq 8.2 \, (\Lambda
R)^{-3/2}
  \end{equation}
while in the second
\begin{equation}
  \label{eq:cut-4D-NDA}
 \widehat \lambda_t(\Lambda) \simeq 4 \, \pi
\left(\frac{1}{ \Lambda R} \right)^{3/2} \simeq 8.9 \, (\Lambda
R)^{-3/2} 
\end{equation}
where  $\widehat \lambda_t=\lambda_t/(2 \pi R)^{3/2}\simeq m_t/v$, so
that $\Lambda R \sim 5$.

In the QLTM model since the 3rd generation is localized towards $y=0$ 
the Yukawa coupling of both the top and the bottom are relevant for
perturbativity and are indeed equal at $MR\simeq4/\pi$. The relation
between $\widehat \lambda_{t,b}$ and $\Lambda$ now depends on the
localization parameter $MR$. This dependence comes from two sources:
the most important one is the relation between the 5D and the 4D
couplings, the other one is the fact that the KK spectrum is shifted in
presence of non-zero masses. Both sources are important for big masses
and the second one is particularly relevant for low
cutoffs. The impact of these two effects of perturbativity can be
easily estimated by means of 4D NDA, remembering that the masses of
the excited states
go like $m^2=(2n/R)^2+M^2$ for large $MR$ and that the relation between the
5D and 4D couplings reads
\begin{equation}
  \label{eq:cutoff_}
\widehat \lambda_{t,b}(\Lambda_{t,b})\simeq
    \frac{m_{t,b}}{v}\frac{2}{\pi M R (\coth(M R \pi/2)\pm 1)}
\end{equation}
with the sign $+$ for the top and the $-$ for the bottom.

\begin{figure}[t]
  \centering
  \includegraphics[width=10cm]{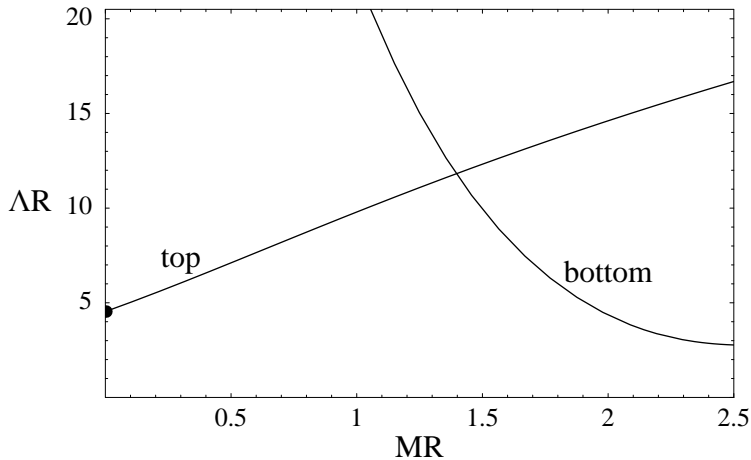}
  \caption{Cutoff estimation from top and bottom Yukawa coupling
  perturbativity as a function of $M R$. The thick point represents the
  case of the CSM.}
  \label{fig:cutoff}
\end{figure}

The estimated cutoff is shown in Fig. \ref{fig:cutoff}. For values of $MR \lesssim 1.4$ the
cutoff is determined by the top Yukawa coupling and increases with
$MR$, while for bigger values $\Lambda R$ is controlled by the
bottom and decreases exponentially with the localization parameter.
In particular for the QLTM one
has successful EWSB for $MR\lesssim 2.3$; at that maximum value of
$MR$ one has $\Lambda R \simeq 3\div4$.  
If the model is not UV completed before $\Lambda$, a
strong interacting sector appears at one of the boundaries. For $\Lambda R \gg 1$ we
can think of the theory in this regime as composed by a bulk
sector with perturbative gauge couplings coupled to a strong
interacting sector localized at one of the boundary. This motivates
the possibility, when considering the effects of higher order SUSY
operators on EWPT, to focus on those localized at the boundaries.

\section{EWPT constraints}
\label{sec:EWPT}

The effects on EWPT arise from various sources. First of all
there are calculable loop effects generated by non
supersymmetric operators. In fact, as we shown in Sect. \ref{sec:model},
supersymmetry is globally broken: the mass splitting between particles belonging
to the same supersymmetric multiplet is  $1/R$. However
the residual supersymmetry of the theory restricts the form of the
possible counterterms. Observables
corresponding to operators which have no counterterm allowed by the
residual supersymmetry are thus necessarily finite.
This is the case for the Higgs boson mass \cite{Barbieri:2000vh}, but
also for several other
observables like $b \to s \gamma$ \cite{Barbieri:2001mr}, the muon
$g-2$ \cite{Cacciapaglia:2001pa} and the Higgs boson
decay into two gluons \cite{Cacciapaglia:2001nz}.
The calculations for $b \to s \gamma$ and $g_\mu -2$ show that for
$1/R \gtrsim 400 \GeV$ these effects are below the current
experimental sensitivity. 

Beyond these calculable effects one should also consider the effect of
supersymmetric operators, mainly those localized at
the boundaries. Indeed one has to look for operators
respecting the $N=1$ supersymmetries present at the boundaries and which
affect EWPT. The fact that they are allowed by supersymmetry makes 1-loop
computations divergent. These operators, whose coefficients have negative
dimension in mass, are generated
by radiative corrections at the scale where perturbation theory breaks
down and are weighted by powers of $1/\Lambda R$. In order to build a fully reliable and predictive theory one has
to consider the impact of such operators on EWPT. We analyze the localized
operators
assuming that their coefficients saturate
perturbation theory at the scale $\Lambda$, estimating them by
5D NDA.

\subsection{Universal effects}
\label{sec:universal}

We make an analysis in terms of the form
factors $\hat S,\hat T,W,Y$ introduced in Ref. \cite{Barbieri:2004qk}. Such an analysis
is valid for a wide class of ``universal theories'' in which the only
gauge interaction (except QCD) of all the
light fermions of the SM is 
\begin{equation}
  \label{eq:Lint}
  \lag_{{\rm int}} = \bar \Psi \gamma^\mu \left(T^a \bar W^a_\mu + Y
  \bar B_\mu \right) \Psi
\end{equation}
where $\bar W, \bar B$ are not necessary the physical $W,B$. In our
case, if no localization parameter is present, it follows from momentum
conservation in the 5th dimension that, at tree level, the ``interpolating'' fields
$\bar W, \bar B$ are exactly the zero modes of the 5D gauge
bosons. We shall come back to the localization effects later on.

Upon use of the equations of motion and neglecting terms vanishing
with the fermion masses, a complete set of dimension-6 operators which
contribute to $\hat S,\hat T,W,Y$  are \cite{Grinstein:1991cd}:
\begin{subequations}
\label{eq:dim6op}
\begin{align}
  \mathcal{O}_{WB} & = \frac{1}{g g'} (H^\dagger \tau^a H) W_{\mu \nu}^a
  B_{\mu \nu} \\
  \mathcal{O}_{H} & = \left|H^\dagger D_\mu H \right|^2 \\
  \mathcal{O}_{BB} & = \frac{1}{2 g'^2} \left( \partial_\rho B_{\mu
  \nu} \right)^2 \\
  \mathcal{O}_{WW} & = \frac{1}{2 g^2} \left(D_\rho W^a_{\mu
  \nu} \right)^2
\end{align}
\end{subequations}
If they appear in a 4D lagrangian as
\begin{equation}
\label{eq:4Dcoeff}
  \delta \lag = \frac{1}{v^2} \left(c_{WB} \mathcal{O}_{WB} + c_{H}
  \mathcal{O}_{H} + c_{WW} \mathcal{O}_{WW} + c_{BB} \mathcal{O}_{BB} \right)
\end{equation}
where $v= \vev{H} = 174 \GeV$ is the Higgs vev, they give the
following contributions to the EW form factors \footnote{The
  normalization of the vector fields is such that
  $\lag_{\rm{kin}}=- \frac{1}{4 g^2} W_{\mu \nu}^a W_{\mu \nu}^a -
  \frac{1}{4 g'^2} B_{\mu \nu} B_{\mu \nu}$}
\begin{equation}
\label{eq:formfactors}
  \hat S = 2 \frac{c_W}{s_W} c_{WB}, \hspace{1cm} \hat T = -c_H,
  \hspace{1cm} W= - g^2 c_{WW}, \hspace{1cm} Y=-g^2 c_{BB}
\end{equation}

Since at the boundaries there are $N=1$ supersymmetries,  we have
to find the supersymmetric completion of
the operators \eq{dim6op}.

This can be easily accomplished using supersymmetric gauge covariant
derivatives (see for example \cite{Gates:1983nr}). Their properties are briefly recalled in appendix
\ref{app:cov-derivatives}.
A simple power counting shows that all the SUSY and gauge invariant
operators up to dimension-6 only, involving Higgs and vector
superfields are
 \begin{subequations}
\label{eq:susyop}
 \begin{align}
 & \int d^4\theta (\hat H^{\dagger} e^{g V} \hat H)^2 \label{eq:susyop1} \\
 & \int d^2\theta \ {\rm Tr} \, \left(\nabla^{\mu} \hat W^{\alpha} \nabla_{\mu} \hat
 W_{\alpha}\right) + \ {\rm h.c.}\label{eq:susyop2}\\
 & \int d^2\theta  \ {\rm Tr} \, \left(C_{\dot \alpha \dot \beta} \nabla^{\alpha \dot
   \alpha}\hat W_{\alpha}\nabla^{\beta \dot \beta}\hat W_{\beta}\right)  + \ {\rm h.c.}
 \label{eq:susyop3}\\
 & \int d^4\theta \hat H^{\dagger} e^{g V} \hat W^{\alpha}  e^{-g V}
 \nabla_{ \alpha} e^{g V} \hat H + \ {\rm h.c.}\label{eq:susyop4} \\
 & \int d^4\theta \hat H^{\dagger} e^{g V} \nabla^{\mu}\nabla_{\mu}
 \hat H \label{eq:susyop5} \\
 & \int d^4\theta \hat H^{\dagger} e^{g V}\hat H \label{eq:susyop6} \\
 &\int d^2\theta  \ {\rm Tr} \, \left(\hat W^{\alpha}\hat W_{\alpha}\right) + \ {\rm h.c.} \label{eq:susyop7}
 \end{align}
 \end{subequations}
where $\hat H$ is
 the Higgs chiral superfield while $V$ is a general
 vector superfield and $\hat W^{\alpha}$ its chiral supersymmetric field strenght . They refers both to the $W$
 and to the $B$ vector
 fields. The symbol $C$ is the usual antisymmetric matrix used in raising and lowering
 spinorial indices. In our notation $C_{\dot\alpha \dot\beta}=
 \sigma_2$. 

Using (anti)commutation rules the other unlisted operators are shown
to be equivalent to suitable combinations of the previous ones.
Expanding them in component fields, one can see that all the operators
in (\ref{eq:dim6op}) are originated, up to terms that vanish by the
equations of motion and/or in the limit of massless
fermions. Therefore the basis selected in \cite{Barbieri:2004qk}
can be supersymmetrized to $N=1$ in 4D. In particular the first
operator originates ${\cal O}_H$, the second and the third originate
${\cal O}_{WW}$ and ${\cal O}_{BB}$ while the fourth
contributes to ${\cal O}_{WB}$. On the contrary the localized kinetic
terms (\ref{eq:susyop}f-g) only contribute through the mixing of the
zero modes with the Kaluza Klein states. The related effects, also of
``universal'' nature, are however subdominant in the parameter region
of interest, with respect to the direct contributions from dimension 6
operators and we shall neglect them in the following.

To estimate the coefficients of the operators \eq{dim6op} we use 5D
NDA. Notice that the existence of the supersymmetric
operators \eq{susyop} makes the 1 loop corrections to these coefficients divergent.
For example, the Yukawa contributions
to $c_H$, or $\hat T$, has a
quadratic sensitivity to the UV, while the dependence of $\hat S$ is
logarithmic. 
This can be easily found by dimensional analysis. In fact the coefficient $c_H$ of the operator ${\cal
  O}_H$ localized at one of the boundaries and written in terms of the
5D fields has dimension of \mbox{(mass)$^{-4}$}.  1-loop Yukawa
corrections will be proportional to the fourth power of the 5D Yukawa
coupling. Then the quadratic dependence on $\Lambda$ comes out
immediately  remembering that the dimension of a 5D
Yukawa coupling is \mbox{(mass)$^{-3/2}$}. In the same way $c_{WB}$
for a localized contribution has dimension of \mbox{(mass)$^{-3}$} and
the Yukawa corrections to ${\cal O}_{WB}$ are proportional to the
square of the Yukawa coupling giving a logarithmic dependence on $\Lambda$.

Let us now suppose that the operators \eq{susyop} contribute to a
localized term $\delta \lag_4$ in eq. \eq{lagrangian}, where the
fields $H,W_\mu^a$ and $B_\mu$ are 5D fields localized at a
boundary. The dominant contribution to the EWPT from the operators of
dimensions 6 in eq. (\ref{eq:susyop}a-e) comes from the zero modes of
the various fields, i.e. the standard gauge and Higgs bosons. 
The localized operators we are interested in are then the
following
\begin{align}
  \delta \lag_4 &= C_H \left|\bar H^\dagger D_\mu \bar H \right|^2 +
  \frac{C_{WB}}{g g'}
  \left(\bar H^\dagger \tau^a \bar H \right) W_{\mu \nu}^a B_{\mu \nu}
  + \frac{C_{BB}}{2 g'^2} \left(
  \partial_\rho B_{\mu
  \nu} \right)^2 + \frac{C_{WW}}{2 g^2} \left(D_\rho W^a_{\mu
  \nu} \right)^2  \label{eq:dim6-5D-1}
\end{align}
where $\bar H$ is the zero mode of the Higgs boson with canonical 5D
normalization, whereas the vector bosons are already normalized to 4D
(and $g,g'$ are the standard 4D gauge couplings). Using naive dimensional
one finds
\begin{align}
\label{eq:5Dcoeff}
  C_H = \frac{(24 \pi^3)^2}{16 \pi^2} \frac{1}{\Lambda^4},
  \hspace{0.8cm} C_{WB} = g g'  \frac{24 \pi^3}{16 \pi^2}
  \frac{1}{\Lambda^3}, \hspace{0.8cm} C_{BB}= C_{WW} = \frac{1}{16 \pi^2}
  \frac{1}{\Lambda^2}
\end{align}
To connect the 5 dimensional coefficients of \eq{5Dcoeff} to the 4
dimensional coefficients of eq. \eq{4Dcoeff} one has to rescale the 5D
Higgs field in terms of the 4D zero mode: $\bar H = H/ \sqrt{2 \pi
  R}$. Using \eq{4Dcoeff} and \eq{formfactors} one gets
\begin{subequations}
\label{eq:corrections}
\begin{align}
 \hat S (\mathcal O_{WB}) & \sim  \frac{3}{2} g^2 \frac{(v
  R)^2}{(\Lambda R)^3}  \simeq 
  1.6 \cdot 10^{-4} \frac{(R \cdot \TeV)^2}{(\Lambda R /5)^3} \\
  \hat T (\mathcal{O}_H) & \sim 9 \pi^2 \frac{(v R)^2}{(\Lambda R)^4}
  \simeq 4.3 \cdot
  10^{-3} \frac{(R \cdot \TeV)^2}{(\Lambda R /5)^4} \label{eq:corrT} \\
  W(\mathcal O_{WW}) & \sim Y(\mathcal O_{BB}) \sim \frac{g^2}{16 \pi^2}
  \frac{(v R)^2}{(\Lambda R)^2} \simeq
  3.3 \cdot 10^{-6} \frac{(R \cdot \TeV)^2}{(\Lambda R/5)^2}
\end{align}
\end{subequations}

\begin{figure}[t]
  \centering
  \includegraphics[width=10cm]{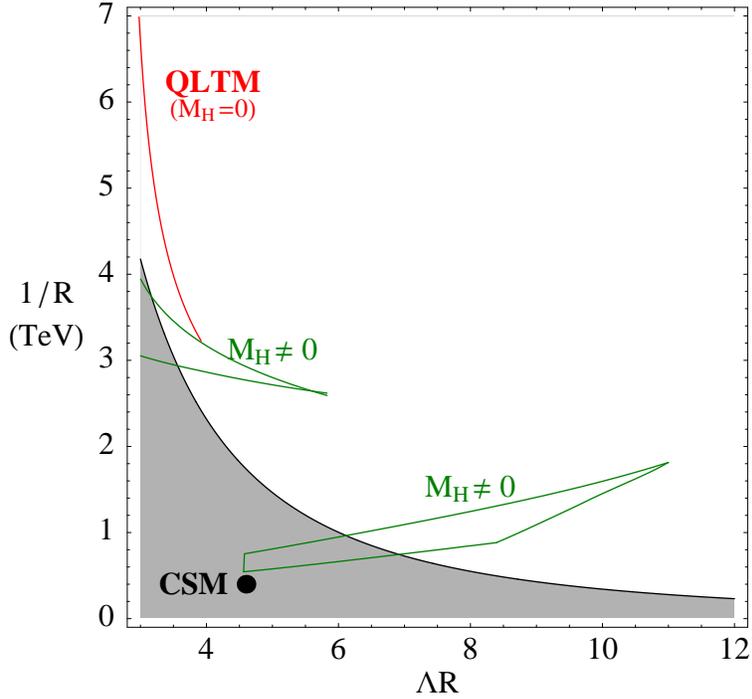}
  \caption{Bounds on the CSM and the QLTM models from EWPT. The shaded
  region is excluded at 99 \% of C.L.}
  \label{fig:EWPT}
\end{figure}

One can notice that the dominant contributions come from
$C_H$. $C_{WW},C_{BB}, C_{WB}$ become comparable to $C_H$ only for
values of the cut-off $\Lambda R \simeq 100$ which can never be
attained (see Fig. \ref{fig:cutoff}). Comparing the
contributions \eq{corrections} to the experimental values \cite{Barbieri:2004qk}
\begin{equation}
  \label{eq:STexp}
  \begin{array}{ll}
\hat S  = (-0.7 \pm 1.3) \cdot 10^{-3} \hspace{1.5cm} & \hat T = (-0.5  \pm 0.9)
 \cdot 10^{-3}, \\
W  = (0.2 \pm 0.6) \cdot 10^{-3}, &  Y  = (0.0 \pm 0.6) \cdot 10^{-3}
  \end{array}
\end{equation}
we can obtain the bound shown in Fig. \ref{fig:EWPT}. The shaded
region is excluded at 99 \% of C.L. The sign of $C_H$ is irrelevant.
The black point represents the
CSM described in Sec. \ref{sec:CSM},  the red continuous line
represents the
QLTM model described in Sec. \ref{sec:QLT} for $M_U=M_Q=M_D$ and $M_H=0$,
while the region inside the green lines corresponds to the QLTM model
varying $M_H$ as described in Sec. \ref{sec:QLT}. This region consists
in two disconnected areas since we excluded the region of the
parameter space where the Higgs is too light and already excluded by
direct searches. The red continuous
line is stopped at $1/R \simeq 3.2 \TeV$ because the Higgs mass drops
below the experimental limit (see Fig. \ref{fig:MHvs1/R}) and at about
7 TeV when a fine-tuning of about 10 \% in the potential
occurs. One has to remember that the connection between $MR$ and $1/R$
is not as precise as the one between the Higgs mass and $1/R$. Thus
the red continuous line can move horizontally  of
about $\Delta (\Lambda R) \simeq 3$ due to this uncertainty. From
these universal effects it seems problematic to reconcile the CSM with
the EWPT. One should not forget, on the other hand, that
eqs. \eq{corrections} are a naive estimate, dependent on a high power
of $\Lambda R$. Furthermore, an unknown contribution coming from the
UV completion of the theory could
provide a fine-tuned cancellation in order to make the size of the
contribution to $\hat T$ in Eq. \eq{corrT}
sufficiently small. Conversely, in the QLTM model both for
$M_H=0$ and for $M_H \neq 0$, the typical values of $1/R$ are large enough
not to create conflicts with EWPT.

\subsection{Non universal effects}
\label{sec:nonuniversal}

We now come back to the non universal effects involving the bottom
quark. 

In the CSM a localized operator
\begin{equation}
  \label{eq:Zbb}
  \int d^4 \theta \; \hat H^\dagger \hat H \; Q_3^\dagger e^{g V} Q_3
\end{equation}
is generated, with a similar coefficient to $C_H$ in eq. \eq{5Dcoeff}. It is a
correction to the $Z b \bar b$ vertex and thus one obtains a similar bound
to the one discussed in the previous Subsection.

In the QLTM and for $M_H=0$ it is the
bottom Yukawa coupling which becomes non perturbative first, hence one
expects that the most important contribution comes from the operator
(\ref{eq:Zbb}) localized at the $y=\pi R/2$
boundary. But since the bottom gets quasi-localized mostly at the $y=0$ boundary, a
wave function suppression is present making this effect
negligible. If we turn $M_H$ on, then successful EWSB occurs for lower
values of $MR$ where it is the top Yukawa coupling
that determines the cut-off. Then one should consider  the operator at
$y=0$. However one obtains a bound that does not differ significantly
from the one obtained above from $\widehat T$, which is weaker then the one coming from
flavor violation effects, analyzed in the following.

In addition to this kind of effect, if we turn on a mass term for the third generation of
quarks, the interpolating field defined by eq. \eq{Lint} is no longer the same for all the
light fermions since, for the bottom, it is a
superposition of all the KK modes. Then for the third generation an
additional interaction to \eq{Lint} is present at the tree level which gives non
universal effects mainly concerning the bottom
quark. These effects produce 4-fermion interactions
involving the bottom quark and modify 
the $Zb\bar b$ vertex only in presence of localized kinetic terms for the
gauge or Higgs multiplets. The size of this effect is comparable to
the one produced by the operator \eq{Zbb}.

The last source of effects comes from flavor violation. If one localizes only the third
quark generation then it is the different coupling of the KK gluons
to the first two generations with respect to the third one which gives
an effect. First of all we have to assume that the Yukawa matrices and
the 5D mass matrices are
diagonal in the same basis. Assuming mixing angles and phases of the down-quark Yukawa
coupling matrix comparable to those of the Cabibbo-Kobayashi-Maskawa
matrix, then the strongest bound comes from the $\epsilon$-parameter
in $K$ physics: one needs $1/R \gtrsim 1.5 \TeV$
\cite{Delgado:1999sv}. This bound applies to the QLTM model. In the
region where $1/R \lesssim 2 \TeV$ (and $M_H \neq 0$) there is a weaker bound due to the
weaker localization. Even if this bound  is stronger than the one
coming from EWPT when $\Lambda R \gtrsim 10$, the region inside the green lines showed in
Fig. \ref{fig:EWPT} is entirely allowed.

We can then conclude that while in the CSM it appears problematic to
avoid a conflict with EWPT without invoking a fine-tuning mechanism,
the QLTM model, which foresees natural
values of $1/R \gtrsim 2 \TeV$, is perfectly compatible with EWPT.

\section{Conclusions}
\label{sec:conclusions}

We have analyzed in detail the constraints coming from EWPT to the
possible implementations of the idea to relate the ElectroWeak scale
to the radius of a compact extra dimension. We have recalled how,
insisting on supersymmetry in 5D and breaking it through boundary
conditions $\grave{\textrm{a}}$ la Scherk-Schwarz, it is possible to
build essentially two kind of models. In the first one, the CSM, the
radius of the extra dimension turns out to be
in the 400 GeV range, while the Higgs mass is about 130 GeV. In the
other possible implementation of the general setup, the QLTM model, we considered a
theory where the third generation of quarks is localized close to the
boundary $y=0$ through a mass term which is of the order of $1/R$. In
this theory the typical value of $1/R$ is increased by almost one order
of magnitude with respect to the CSM due to a quasi cancellation
between the electroweak and the top corrections to the Higgs mass. We
analyzed both the case of vanishing Higgs mass term and the case where
$M_H \neq 0$.

The possible effects come from:
\begin{enumerate}
\item \label{calculable} Calculable loop effects generated by non supersymmetric
  operators.
\item \label{localized} Supersymmetric operators localized at the boundaries.
\item \label{flavor} Flavor physics.
\end{enumerate}
We have shown that the strongest constraints come from \ref{localized}
and \ref{flavor}. In particular we have shown that it is difficult to
reconcile the CSM with \ref{localized} without invoking some
fine-tuning mechanism which suppresses the effects of the higher order
operators to the $\rho$ parameter ($\hat T$) or to the $Z b \bar b$
vertex.

On the contrary we have shown that the QLTM model, both with $M_H=0$ and
$M_H \neq 0$, have no problem with EWPT. The bound \ref{localized} is
stronger for low values of the cut-off: for $\Lambda R \gtrsim 10$ the
bound \ref{flavor} dominates. However the regions corresponding to
the QLTM are always allowed.

We can then conclude that the idea of supersymmetry breaking by
boundary conditions along an extra dimension successfully addresses
the Little Hierarchy problem. It allows to build very predictive models with a
naturally light Higgs boson, insensitive to the UV. The phenomenology
of such models is very rich and testable at the
Tevatron and the LHC. We have shown that at least one of the possible
implementations of this idea is perfectly compatible with all the
possible constraints coming from EWPT.

\bigskip

\paragraph{Acknowledgments}
We warmly thank Riccardo Barbieri for providing us important insights
to this work. We also thank Riccardo Rattazzi andAlessandro Strumia
for useful discussions. This work was supported in part by MIUR and
by the EU under TMR contract
HPRN-CT-2000-00148.

\appendix
\section{Supersymmetric gauge covariant derivatives}
\label{app:cov-derivatives}

The supersymmetric  gauge covariant derivatives in the chiral
representation are defined as
\begin{equation}
  \label{eq:cov-derivatives}
\nabla_{\alpha}= e^{- g V} D_{\alpha} e^{g V} \hspace{.5cm} \bar \nabla_{\dot \alpha} =
 \bar D_{\dot \alpha}  \hspace{.5cm} \nabla_{\alpha
  \dot \alpha}=-i \{\nabla_{\alpha}, \bar \nabla_{\dot \alpha}\}  
\end{equation}
where $D_{\alpha}$, $\bar D_{\dot \alpha}$ are ordinary SUSY covariant
chiral derivatives.
Under a gauge transformation they transform as 
\begin{equation}
  \label{eq:cov-transf}
  \nabla_A \rightarrow e^{\Lambda} \nabla_A e^{-\Lambda}
\end{equation}
where $A=\alpha,\dot \alpha,\mu$.
In this basis, chiral and antichiral superfields transform as
\begin{equation}
  \label{eq:field-transf}
\Phi \rightarrow e^{\Lambda} \Phi \hspace{.5cm} \bar \Phi e^{g V}
\rightarrow \bar \Phi e^{g V} e^{-\Lambda}
\end{equation}
under a gauge transformation.

\end{document}